\documentclass[fleqn,usenatbib]{mnras}

\usepackage{newtxtext,newtxmath}
\usepackage[T1]{fontenc}

\DeclareRobustCommand{\VAN}[3]{#2}
\let\VANthebibliography\thebibliography
\def\thebibliography{\DeclareRobustCommand{\VAN}[3]{##3}\VANthebibliography}

\usepackage{graphicx}	% Including figure files
\usepackage{amsmath}	% Advanced maths commands
\usepackage{lipsum}
\usepackage{pdflscape}
\title[UDG Masses in Simulations and Observations]{Ultra-Diffuse, Ultra-Different: Observed vs. Simulated Ultra-Diffuse Galaxies Live in Fundamentally Different Halos}

\author[J. S. Gannon et al.]{Jonah S. Gannon$^{1}$,\thanks{E-mail: jonah.gannon@gmail.com (JSG)}
Arianna Di Cintio$^{2,3}$,
Duncan A. Forbes$^{1}$,
Guacimara Garc\'ia-Bethencourt$^{2}$,
\newauthor
Jean P Brodie$^{1,4}$,
Noam Libeskind$^{5}$,
Warrick J. Couch$^{1}$
and
Johanna Hartke$^{6,7,8}$
\\
$^{1}$Centre for Astrophysics \& Supercomputing, Swinburne University of Technology, Hawthorn VIC 3122, Australia\\
$^{2}$Departamento de Astrof\'isica, Universidad de La Laguna, E-38200, La Laguna, Tenerife, Spain\\
$^{3}$Instituto de Astrof\'isica de Canarias, Av. Via Lactea s/n, E38205 La Laguna, Spain\\
$^{4}$Department of Astronomy \& Astrophysics, University of California Santa Cruz, 1156 High Street, Santa Cruz, CA 95064, USA\\
$^{5}$ Leibniz-Institut fur Astrophysik Potsdam; An der Sternwarte 16, 14482 Potsdam, Germany \\
$^{6}$Finnish Centre for Astronomy with ESO, (FINCA), University of Turku, FI-20014 Turku, Finland\\
$^{7}$Tuorla Observatory, Department of Physics and Astronomy, University of Turku, FI-20014 Turku, Finland\\
$^{8}$ Turku Collegium for Science, Medicine and Technology (TCSMT), University of Turku, FI-20014 Turku, Finland
}

\date{Accepted XXX. Received YYY; in original form ZZZ}

\pubyear{\the\year{}}

\begin{document}
\label{firstpage}
\pagerange{\pageref{firstpage}--\pageref{lastpage}}
\maketitle

\begin{abstract}
In this work, we compare galaxies from the NIHAO and HESTIA simulation suites to ultra-diffuse galaxies (UDGs) with spectroscopically measured dynamical masses. For each observed UDG, we identify the simulated dark matter halo that best matches its dynamical mass. In general, observed UDGs are matched to simulated galaxies with lower stellar masses than they are observed to have. These simulated galaxies also have halo masses much less than would be expected given the observed UDG's stellar mass and the stellar mass -- halo mass relationship. We use the recently established relation between globular cluster (GC) number and halo mass, which has been shown to be applicable to UDGs, to better constrain their observed halo masses. This method indicates that observed UDGs reside in relatively massive dark matter halos. This creates a striking discrepancy: the simulated UDGs are matched to the dynamical masses of observed ones, but not their total halo masses. In other words, simulations can produce UDGs in halos with the correct inner dynamics, but not with the massive halos implied by GC counts. We explore several possible explanations for this tension, from both the observational and theoretical sides. We propose that the most likely resolution is that observed UDGs may have fundamentally different dark matter halo profiles than those produced in NIHAO and HESTIA. This highlights the need for a simulation that self-consistently produces galaxies of a stellar mass of $\sim 10^8 M_\odot$ in dark matter halos that exhibit the full range of large dark matter cores to cuspy NFW-like halos.
\end{abstract}


%%%%%%%%%%%%%%%%%%%%%%%%%%%%%%%%%%%%%%%%%%%%%%%%%%

%%%%%%%%%%%%%%%%% BODY OF PAPER %%%%%%%%%%%%%%%%%%

\section{Introduction}

UDGs are characterised by low surface brightness and large size. In particular, \citet{vanDokkum2015} assigned a working definition of central g band $\mu$ $\ge$ 24 mag. arcsec$^{-2}$ and effective radius R$_e$ $\ge$ 1.5 kpc. This corresponds to dwarf galaxy-like stellar masses. These selection criteria are continuous, rather than discrete, from known galaxies with slightly higher surface brightness and smaller sizes. Galaxies that nudge up against the UDG criteria have been referred to as NUDGes \citep{Forbes2024}. There is no doubt that there are many different evolutionary pathways for a galaxy to occupy the parameter space assigned to UDGs (see e.g., \citealp{FerreMateu2023, Buzzo2025}). 

Perhaps more interesting is when the properties of UDGs are extreme when compared to dwarf galaxies of a similar stellar mass. These properties include their globular cluster (GC) systems and their halo masses. Several studies have found high GC numbers (or system mass) to stellar mass ratios \citep{vanDokkum2017, Lim2018, Forbes2020, Danieli2022, Saifollahi2022}. Halo masses are very difficult to measure directly, with only one UDG, DF44, having a halo mass estimate available based on its radially-extended kinematics. In this case a massive halo, for its stellar mass, was indicated. However, as with most mass modelling, the halo mass is subject to caveats on the shape of the mass profile and orbital anisotropy \citep{vanDokkum2019b, Wasserman2020}.

\citet{Forbes2024} derived the halo mass for UDGs with more than 20 GCs using two different methods. The first method used the empirical scaling between GC count and halo mass \citep{Burkert2020}. The second method estimated the total halo mass based on the enclosed dynamical mass derived from measured velocity dispersions. The latter method required an assumption of a mass profile. Here they explored both an NFW cusp \citep{NFW} and a core (e.g., \citealp{Read2016}). The core was further assumed to be `maximal' and equal to 2.75 times the observed half-light radius following \citet{Read2016}. The relation between halo concentration and halo mass of \citet{Dutton2014} was also followed. \citet{Forbes2024} concluded that halo masses derived from cored mass profiles were in better agreement with GC-inferred halo masses than cusp profiles. The halo masses inferred for these GC-rich UDGs were over-massive compared to standard stellar mass--halo relations (SMHRs), suggesting that such galaxies had failed to form stars in the expected amount. These `failed galaxies' appear to challenge standard models of galaxy formation. We note however that the initial idea of `Failed $L_{*}$ galaxies' \citep{vanDokkum2015} has been largely rejected, as the halo masses of the galaxies do not reach into the $L_{*}$ regime, despite being larger than is standard for a dwarf \citep{Sifon2018, Gannon2020}. Dwarf galaxies of a similar stellar mass with high halo masses have also been dubbed `baryon deficient' \citep{ManceraPina2025}. 

The approach of \citet{Forbes2024} required various assumptions to infer the halo masses, including the unknown size of the core and the halo concentration parameter. An alternative approach is to use UDG models generated from cosmological simulations. This has the advantage that the galaxy properties are `built-in' based on the physics included in the simulations. Comparisons to these simulations can then help test the model of the physics included. Galaxies matching the UDG criteria have naturally arisen in various simulations (e.g., NIHAO; \citealp{DiCintio2017,Jiang2019, CardonaBarrero2020, CardonaBarrero2022}, HESTIA; \citealp{Newton2023}, FIRE; \citealp{Chan2018}, the Illustris Suite; \citealp{Carleton2019, Sales2020, Doppel2021, Benavides2021, Benavides2023}, Romulus; \citealp{Tremmel2020, Wright2021} and MAGNETICUM; Gannon et al. Submitted, to name a few). These simulations span the full range of environments from the low-density field to massive, dense galaxy clusters. Indeed, many simulations have been found to reproduce the dynamical masses of UDGs, which has been presented as evidence that they are producing realistic UDGs \citep{DiCintio2017, Chan2018}. Further research has been conducted contrasting simulations and the full range of UDG properties, for example, galaxy sizes and HI content \citep{DiCintio2017}, and their radial distribution within the Local Group \citep{Newton2023}. However, these studies are typically conducted at the population level, rather than through direct galaxy-by-galaxy comparisons between simulations and observations. Moreover, only recently has it become possible to perform detailed studies involving resolved observational quantities, such as stellar populations and metallicity gradients (e.g., \citealt{KadoFong2022b, FerreMateu2023, CardonaBarrero2022,  Villaume2022}).

\begin{figure*}
    \centering
    \includegraphics[width=0.95\linewidth]{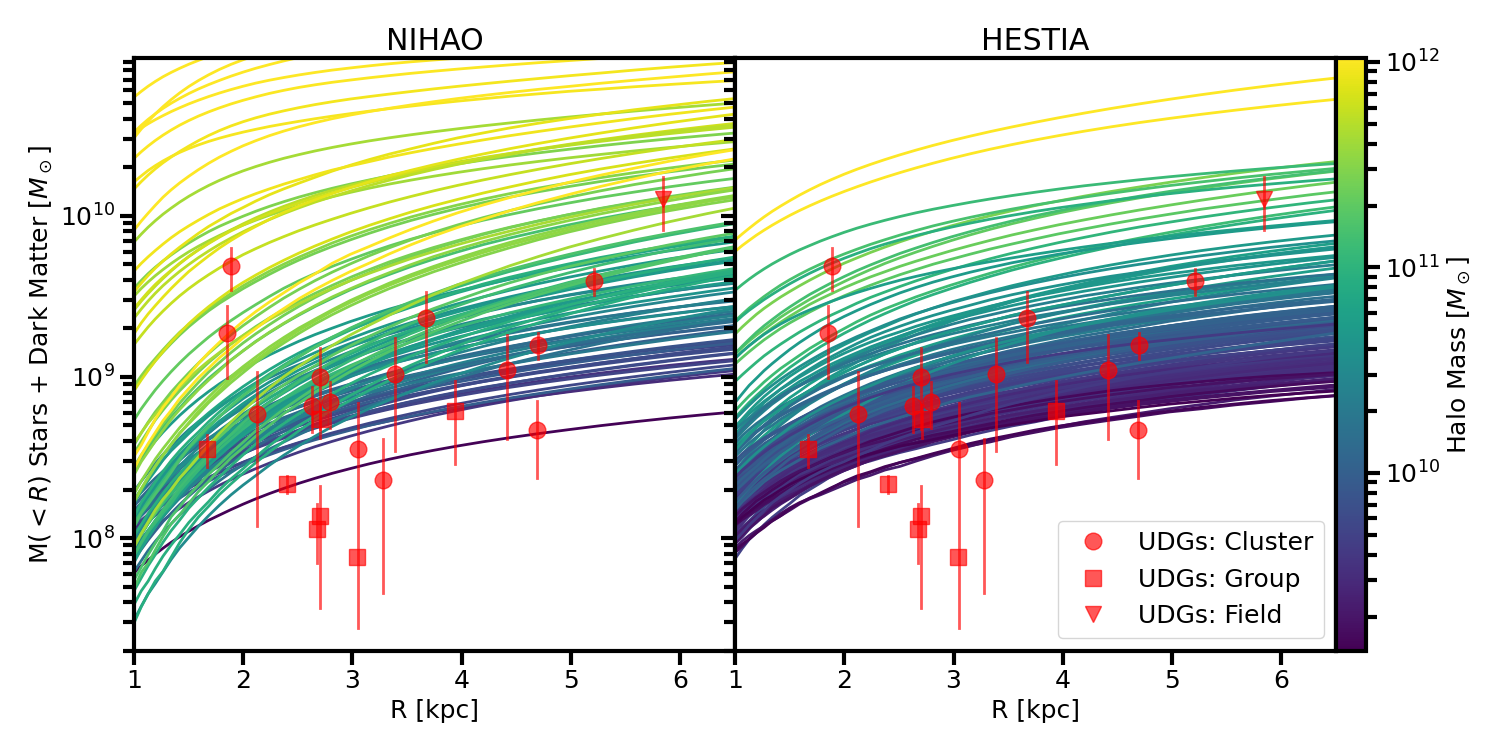}
    \caption{Dynamical mass enclosed within a radius vs that radius. Red points are observed UDGs with markers corresponding to their being in a field (triangle), group (squares) or cluster (circles) environment. On the \textit{left} we compare the observed dynamical masses to mass profiles from the NIHAO simulation, and on the \textit{right} we compare the observed dynamical masses to mass profiles obtained from the HESTIA simulation. For both simulations, their mass profiles are colour-coded by their total halo mass, and we do not limit the plotted halos to just those that contain UDGs. For both simulations, we exclude any gas mass when plotting, as the observed UDGs are largely gas-free. UDGs are matched to the halo that best matches their observed dynamical mass for comparison in Figure \ref{fig:smhm}. It is of note that while many UDG dynamical masses prefer massive halos ($ >10^{11}~M_{\odot}$), none reside in halos as massive as the Milky Way ($\sim 10^{12}~M_\odot$).
    % {\it DF: is there something odd about the lowest profile? No UDG lives in a MW-like halo}
    }
    \label{fig:mass_radius}
\end{figure*}

In this work, we take a further step and compare the individual dark matter halos in simulations that best reproduce observed UDG dynamical masses to the individual observed galaxies they match. We emphasise a comparison of the stellar mass forming within these simulated halos and the total mass of these halos in comparison to the observed total halo masses of the UDGs they match. Section \ref{sec:data} presents the simulated and observed data we use in this work. Section \ref{sec:methods} describes the method of matching simulations to observations and how we compare the stellar mass in the simulation to the stellar mass of the observed galaxy. In Section \ref{sec:discussion}, we discuss these results. We place particular emphasis on the UDG's positioning in stellar mass -- halo mass space within the simulation and on the comparative differences between the simulated and observed halo masses. Finally, we discuss possible causes for the difference between what is seen in observations and simulations. The conclusions of our study are summarised in Section \ref{sec:conclusions}.

% \begin{landscape}
    \begin{table*}
        \centering
        \resizebox{\textwidth}{!}{%
        \begin{tabular}{ccc|ccc|ccc}
        \hline
        \multicolumn{3}{l|}{Observations} & \multicolumn3{c|}{NIHAO} & \multicolumn{3}{c}{HESTIA} \\ \hline
        Name & $M{_{\rm \star, Obs.}}$ & $M_{\rm Dyn.}$ & $M_{\rm Halo, NIHAO}$ & $M_{\rm \star, NIHAO}$ & $F_{\rm \star, NIHAO}$ & $M_{\rm Halo, HESTIA}$ & $M_{\rm \star, HESTIA}$ & $F_{\rm \star, HESTIA}$ \\
         & $[\times 10^{8} M_{\odot} ]$  & $[\times 10^{8} M_{\odot} ]$ & $[\times 10^{10} M_{\odot} ]$ & $[\times 10^{8} M_{\odot} ]$ & $[dex]$ & $[\times 10^{10} M_{\odot} ]$ & $[\times 10^{8} M_{\odot} ]$ & $[dex]$ \\ \hline
        Andromeda XIX & 0.016 & 1.14 & 0.14  (nan, nan) & 0.0 & inf & 0.22 ( nan, nan) & 0.01 & 0.17 \\
        Antlia II & 0.017 & 0.76 & 0.14 (nan, nan) & 0.0 & inf & 0.22 (nan, nan) & 0.01 & 0.2 \\
        DF 44 & 3.0 & 39.54 & 4.29 (2.7, 16.8) & 0.95 & 0.5 & 4.99 (1.73, 5.5) & 8.3 & -0.44 \\
        DFX1 & 3.4 & 23.07 & 4.29 (1.29, 42.33) & 0.95 & 0.55 & 2.46 (0.42, 5.9) & 7.08 & -0.32 \\
        DGSAT-I & 4.0 & 127.85 & 43.29 (14.75, 43.29) & 43.97 & -1.04 & 19.04 (5.08, 22.79) & 53.58 & -1.13 \\
        Hydra-I UDG 11 & 0.63 & 5.92 & 3.26 (0.14, 42.33) & 0.3 & 0.33 & 0.82 (0.14, 5.9) & 0.28 & 0.35 \\
        Hydra-I UDG 12 & 1.19 & 18.69 & 116.47 (14.75, 117.55) & 188.38 & -2.2 & 12.88 (1.73, 12.88) & 38.25 & -1.51 \\
        Hydra-I UDG 4 & 10.6 & 3.59 & 0.14 (0.14, 3.3) & 0.0 & inf & 0.22 (0.14, 0.86) & 0.01 & 2.99 \\
        Hydra-I UDG 7 & 0.49 & 49.08 & 57.47 (57.47, 68.85) & 147.7 & -2.48 & 9.2 (9.2, 27.49) & 47.27 & -1.98 \\
        Hydra-I UDG 9 & 1.78 & 10.43 & 2.13 (0.31, 16.8) & 0.13 & 1.14 & 1.37 (0.14, 3.86) & 0.5 & 0.55 \\
        NGC 1052-DF2 & 2.0 & 1.36 & 0.14 (nan, nan) & 0.0 & inf & 0.22 (nan, nan) & 0.01 & 2.27 \\
        NGC 5846\_UDG1 & 1.1 & 5.46 & 0.6 (0.31, 8.93) & 0.0 & 2.42 & 0.44 (0.16, 1.14) & 0.06 & 1.3 \\
        PUDG\_R15 & 2.59 & 2.29 & 0.14 (0.14, 0.14) & 0.0 & inf & 0.18 (0.15, 0.22) & 0.0 & 4.0 \\
        PUDG\_R16 & 5.75 & 4.71 & 0.14 (0.14, 0.14) & 0.0 & inf & 0.18 (0.14, 0.43) & 0.0 & 4.34 \\
        PUDG\_R84 & 2.2 & 6.61 & 9.28 (0.36, 16.8) & 4.93 & -0.35 & 0.48 (0.16, 1.37) & 0.14 & 1.21 \\
        PUDG\_S74 & 7.85 & 15.86 & 1.4 (0.86, 3.54) & 0.03 & 2.36 & 1.05 (0.42, 1.99) & 0.81 & 0.99 \\
        Sagittarius dSph & 1.32 & 2.18 & 0.14 (0.14, 0.14) & 0.0 & inf & 0.22 (nan, nan) & 0.01 & 2.09 \\
        UDG1137+16 & 1.4 & 6.18 & 0.39 (0.14, 0.95) & 0.0 & 3.66 & 0.26 (0.14, 0.86) & 0.01 & 2.36 \\
        VCC 1287 & 2.0 & 11.11 & 0.65 (0.14, 3.54) & 0.02 & 2.06 & 0.53 (0.14, 1.99) & 0.16 & 1.1 \\
        WLM & 0.41 & 3.56 & 1.52 (0.51, 42.33) & 0.09 & 0.66 & 0.53 (0.3, 3.72) & 0.16 & 0.41 \\
        Yagi275 & 0.94 & 9.99 & 11.79 (0.36, 42.33) & 5.76 & -0.79 & 0.84 (0.16, 5.9) & 0.61 & 0.19 \\
        Yagi358 & 1.38 & 7.03 & 1.46 (0.36, 16.8) & 0.06 & 1.36 & 0.55 (0.16, 1.37) & 0.17 & 0.92 \\
        \hline
        \end{tabular}}
        \caption{A summary of the main results of this work. From left to right columns are: 1) Observed UDG name, 2) Observed UDG stellar mass, 3) Observed UDG enclosed dynamical mass, 4) The total mass of best fitting halo from NIHAO with brackets indicating (the minimum halo mass from NIHAO which passes within the uncertainty on the dynamical mass, the maximum halo mass from NIHAO which passes within the uncertainty on the dynamical mass), 5) The stellar mass of that halo in NIHAO and 6) The logarithmic difference between the observed stellar mass and that which has been simulated in NIHAO (see Equation \ref{eqtn:frac}). Columns 7-9 are the same as columns 4-6, but compare to the HESTIA simulation instead of NIHAO. When infinite values (inf) are listed for the logarithmic difference, it is due to the galaxy being assigned to a `dark halo' in NIHAO (i.e., a dark matter halo that did not form stars). nan values are listed as uncertainties when no halo passes through the dynamical mass calculated. In these cases, the dynamical masses are simply lower than the simulations produce for this halo mass range. }
    
        % Many of the observed UDGs have strongly different halo masses from the galaxies that have formed in a similar halo in NIHAO.
        \label{tab:summary}
    \end{table*}
% \end{landscape}

\section{Data} \label{sec:data}
\begin{figure*}
    \centering
    \includegraphics[width=0.95\linewidth]{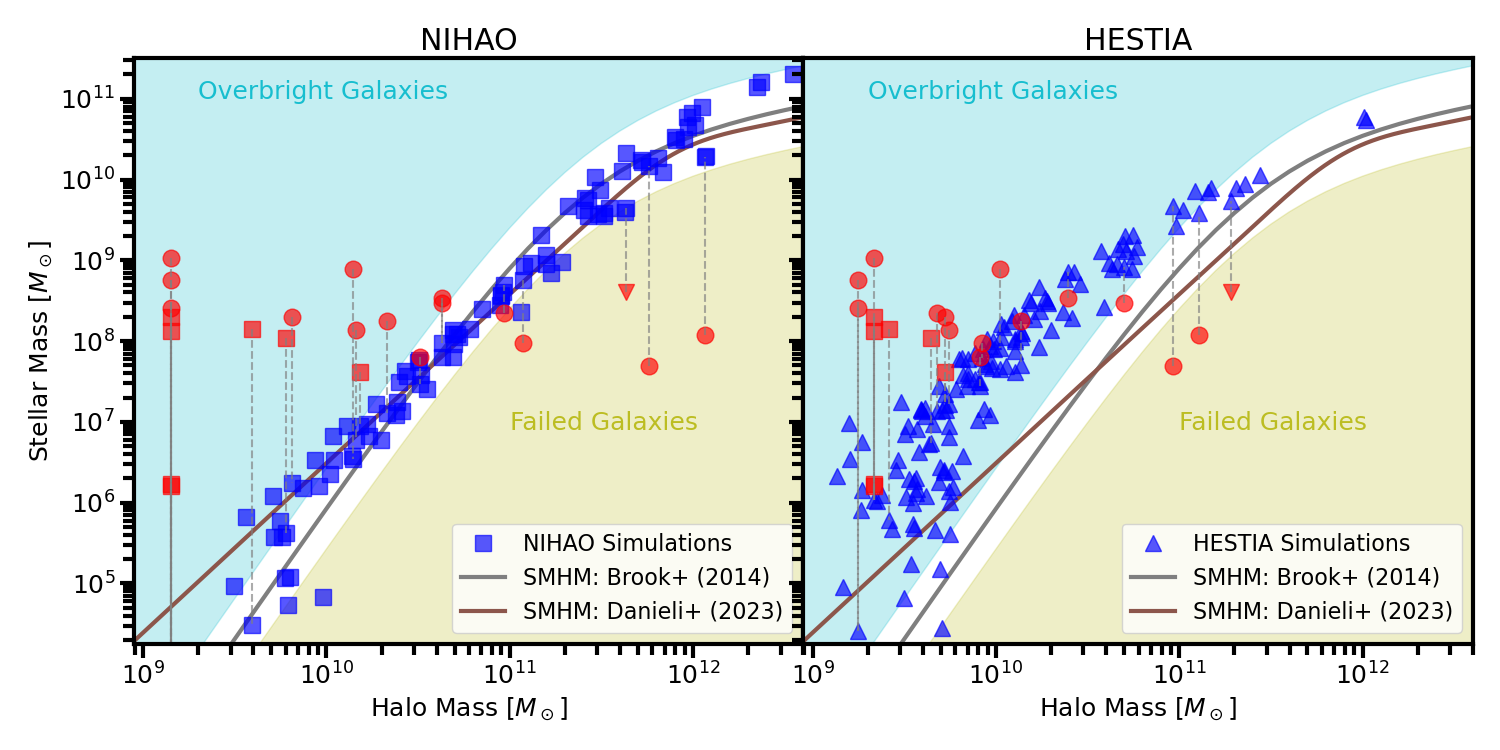}
    \caption{The stellar mass -- halo mass relationship. On the \textit{left} we show galaxies in the NIHAO simulation (blue squares). On the \textit{right} we show galaxies in the HESTIA simulation (blue triangles). We plot UDGs using their observed stellar masses and the total halo mass from their best-fitting dark matter halo after matching them with simulations, as shown in Figure \ref{fig:mass_radius} (red points, with markers per Figure \ref{fig:mass_radius}). We connect the observed UDG to its best-fitting simulated galaxy using a dotted grey line. On the left of the plot, there are 7 UDGs that are connected to a NIHAO dark matter halo that did not form any stellar mass (possibly due to reionisation, a known effect at such low halo masses). As such, their dotted grey lines overlap and connect to a datum outside the plotted region. In both panels, we include an observed stellar mass -- halo mass relationship for normal galaxies from \citet[][grey solid line]{Brook2014} and \citet[][brown solid line]{Danieli2023}. We define regions of >0.5 dex beyond the relationship of \citet{Brook2014} and label them as the regions of overbright galaxies (cyan; i.e., where galaxies have more stellar mass than expected given their total halo mass) and of failed galaxies (olive; i.e., where galaxies have less stellar mass than expected given their total halo mass). HESTIA tends to create galaxies that lie above this relationship, which suggests that their halos over-produce stars (i.e., the simulation suffers from over-cooling). NIHAO creates galaxies that largely follow the established relationship. In both simulations, observed UDGs tend to have best-fitting halos for their dynamical masses which host galaxies of markedly different (usually lower) stellar mass in the simulation. We take this as observational evidence for the need for increased scatter in the stellar mass -- halo mass relationship in the stellar mass range of dwarf galaxies ($M_\star \approx 10^8$ M$_\odot$). }
    % {\it DF: so for models above Mhalo $\sim$ 8x10$^9$ HESTIA does not produce normal galaxies - is this a problem?. Also Moster is an extrapolation, consider Brook or Shany's SMHR.}

    \label{fig:smhm}
\end{figure*}

\subsection{Simulation data}
This paper uses simulations of galaxies from the NIHAO \citep{wang15} and HESTIA \citep{libeskind_hestia_2020} projects. These two simulation suites both produce a population of UDGs, but in different environments and through distinct formation mechanisms. In each simulation, we select haloes with masses in the range $M_{\mathrm{halo}} \sim 10^{9} - 10^{12}\, M_\odot$, ensuring they are not satellites of larger systems (i.e., only isolated galaxies are considered, even in the Local Group environment). In both simulation sets, haloes are identified using the AHF halo finder \citep{knollmann_ahf_2009}.\\

% Additionally, we require that each selected halo contains at least one stellar and one gas particle to exclude completely dark objects.

% This allows for a more comprehensive and fair comparison with observational data.

The NIHAO project includes high-resolution simulations of isolated galaxies, evolved using the SPH code Gasoline \citep{wadsley04}. The code includes subgrid models for metal and energy mixing, UV heating, ionization, and metal-line cooling \citep{shen10}. Star formation and feedback follow the framework of previous MaGICC simulations \citep{stinson13}, which reproduces key galaxy scaling relations \citep{brook12b}, with a star formation threshold of $n_{\mathrm{th}} = 10.3\,\mathrm{cm}^{-3}$. Feedback includes both supernovae \citep{keller14} and early stellar radiation \citep{stinson13}. High resolution ensures that half-light radii are well resolved across a broad mass range. Specifically, in NIHAO particle masses are chosen to ensure that each halo includes $\sim 10^6$~dark matter particles and force softenings are chosen to be $\sim 0.3\%$ of the virial radius \citep{wang15}, ensuring that the half-light radii are well resolved. The NIHAO sample includes central, isolated galaxies from dwarf to Milky Way mass, matching abundance-matching predictions and showing realistic stellar, gas, and dark matter properties \citep{tollet15,maccio16}.  The NIHAO galaxy suite provided the first simulated formation scenario for UDGs within a $\Lambda$CDM framework \citep{DiCintio2017}, and has since been widely used to investigate various UDG properties in connection with observations. These include studies of stellar metallicity gradients \citep{CardonaBarrero2022} and the degree of rotational support \citep{CardonaBarrero2020}.

Unlike NIHAO, which focuses on isolated galaxies, the HESTIA simulation suite models the formation and evolution of galaxies within a realistic Local Group environment in a fully self-consistent manner \citep{libeskind_hestia_2020}. It employs the moving-mesh code AREPO \citep{weinberger_arepo_2020} along with the AURIGA galaxy formation model \citep{grand_auriga_2017}, which takes into account the most important physical processes relevant for the formation and evolution of galaxies. It includes cooling of gas via primordial and metal cooling, a spatially uniform UV background, star formation via a gas density threshold of 0.13 cm$^{-3}$, stellar and AGN feedback, as well as the implementation of magnetic fields.

Using observationally constrained estimates of the peculiar velocity field \citep{tully_cosmicflows-2_2013}, the initial conditions of the HESTIA simulations are designed to reproduce the main gravitational features of the Local Group's surroundings \citep{hoffman_constrained_1991}. As a result, the simulated Local Group analogues at $z=0$~are embedded within a large-scale structure that closely matches the observed cosmic environment. The high-resolution HESTIA simulation used here consists of two overlapping spherical volumes with radii of 2.5 $h^{-1}$ Mpc, each centred on the Milky Way and M31 analogs at z = 0. This run is labelled 37\_11. The spatial resolution achieved is 177 pc, and the effective masses of the dark matter and gas particles are $M_{\rm DM} = 2 \times 10^5 M_\odot$~and $M_{\rm Gas} = 2.2 \times 10^4 M_\odot$, respectively. The results are robust across all three high-resolution realisations. HESTIA has recently been used to study the properties of UDGs in Local Group-like environments, revealing the presence of a diffuse galaxy population in the simulations that may yet be uncovered observationally \citep{Newton2023}.\\

The formation mechanism of UDGs in the HESTIA simulations is different from that proposed in NIHAO. In NIHAO, UDG formation is driven by repeated gas outflows triggered by supernova (SN) explosions, which in turn reduce both the dark matter and central stellar densities in haloes with $M_{\mathrm{halo}} \sim 10^{10} - 10^{11.5}\ M_\odot$ \citep{DiCintio2017}. This process results in shallower central density profiles (i.e., core-like), as seen in the left panel of Figure \ref{fig:mass_radius}, where many NIHAO galaxies display steeper profiles compared to those in HESTIA over the same mass range. In contrast, HESTIA galaxies retain a cuspy, NFW-like profile across all halo masses. Here, UDG formation is primarily merger-driven: a strong correlation is observed between merger events, a sharp increase in the halo spin parameter, and a sudden rise in effective radius ($R_{\rm e}$). During these events, older stars are dynamically heated and displaced to the galaxy outskirts, while new stars form in extended regions from cold gas accreted during the mergers (Cardona-Barrera et al., in prep.).

\subsection{Observational Data}
% {\it DF: consider moving list of refs to an appendix}\\

The observational data used in the paper come from the catalogue of UDGs with spectroscopic measurements from \citet{Gannon2024}. We retrieved the catalogue on 2025, February 25 when it contained 37 UDGs. The full references for the catalogue are provided in the Data Availability Section. To date, the catalogue is heavily biased towards UDGs in cluster environments, with only a handful of UDGs in the catalogue residing in a group or in the field. As such, the majority of the catalogued UDGs are old and quiescent at present times. 

% Catalogued UDG stellar velocity dispersions usually of order 10-30 km s$^{-1}$, which leads to dynamical masses of order $10^8 -10^{10} M_\odot$ within the half light radius (see e.g., Figure \ref{fig:mass_radius}). 

\section{Methods} \label{sec:methods}
Here we make use of the 22 galaxies in the catalogue with stellar velocity dispersion measurements \citep{Gannon2024}. We use these velocity dispersions, along with their 2D half-light radii to calculate dynamical masses within their 3D, circularised half-light radii using the mass estimator of \citet{Wolf2010}. Uncertainties in our dynamical mass measurements are based solely on the uncertainty in their velocity dispersions, which dominates over the uncertainty in their half-light radii. It is also worth noting that the measurement of velocity dispersion is a good approximation of the second-order velocity moment, which will naturally incorporate any rotation within the same aperture \citep{Courteau2014}.

Of these 22 galaxies, 14 are located in a cluster environment, 7 are in groups (4 of which are from the Local Group), and 1 is in the field. This is of particular note as neither the NIHAO or HESTIA simulations are able to probe the dense cluster environments where the majority of our sample resides. We discuss this as a possible bias to our study in Section \ref{sec:stripping}.

In Figure \ref{fig:mass_radius} we plot the mass profiles from the simulations and overlay our observed data. It should be noted that these simulated mass profiles exclude any gas content, as the observed galaxies are found to be largely gas-free. While we perform this matching for all dark matter haloes in the mass range $10^{9} - 10^{12} M_\odot$, regardless as to whether or not the simulation formed a UDG, we note that the vast majority of the galaxies formed by NIHAO in this mass range are UDGs (see e.g., \citealp{Jiang2019} figure 2). Of particular interest in Figure \ref{fig:mass_radius} is the relative self-similarity of dark matter halos in the simulation. That is, while there may be some variation in the halo profile shape (i.e., core \textit{vs.} cuspiness) over the mass range considered, this variation does not occur at fixed halo mass. Put another way, at fixed halo mass, the simulations do not simultaneously produce both a cusp and a core, i.e., they do not solve the diversity of rotation curves problem \citep{Oman2015}.

In order to match these dynamical masses to their best-fitting dark matter halo in the simulations, we interpolate the simulated halo profiles with a cubic spline and generate a mass at the observed radius for each galaxy. We then assign the observed UDG to the halo that most closely reproduces its observed mass within the half-light radius for comparison. We derive uncertainties on this fit by taking the maximum and minimum halo masses passing through the uncertainties on the dynamical masses. It is of note that we do not make any selection on the galaxies forming within these halos in the simulation. That is, we do not require them to be UDGs. In this way, we select the halo in the simulation that is most similar to the observed halo to allow a comparison with the galaxy that has formed within it. 

We then assign the observed UDG to the simulated galaxy's halo mass and then calculate the logarithmic ratio between the observed stellar mass ($M_{\rm \star, Obs.}$) and the simulated stellar mass within that best fitting halo ($M_{\rm \star, NIHAO}$/$M_{\rm \star, HESTIA}$) as:

\begin{equation} \label{eqtn:frac}
    F_{\rm \star, NIHAO} = \log(\frac{M_{\rm \star, Obs.}}{M_{\rm \star, NIHAO}})
\end{equation}

When $F_\star$~is $<0$~it shows the best fitting dark matter halo in the simulation has formed more stars than the UDG we observe to be similar to it. Conversely, if this value is $>0$~it shows the best-fitting dark matter halo in the simulation has formed fewer stars than the UDG we observe. When this number is $<<0$~(e.g., $-2$) we suggest that these observed galaxies are good examples of what is meant by a ``failed galaxy'', i.e., assuming the total dark matter halo mass that it has been matched to is correct, it has formed far fewer stars than what is expected given simulated galaxy residing in that halo. We calculate an equivalent property $F_{\rm \star, HESTIA}$~for the HESTIA simulation. A summary of derived dynamical masses, best matching halo masses, the stellar masses of those halo masses in their respective simulation and the comparative $F_\star$~values are available in Table \ref{tab:summary}. 

% {\it DF: here F is a lack of stars relative to expected stars, rather than lack of stars wrt expected from its halo mass. }

% note that 1 halo was going to be deleted due to having no stellar mass 1890000000.00 1428402600.16 3050801.62 0.00 1425351798.53
% 6 galaxies, Antlia II, Hydra UDG 4, DF2, R15, R16, Dsph

\section{Discussion} \label{sec:discussion}
\subsection{The Stellar Mass -- Halo Mass Relationship of UDGs} \label{sec:simulated_hm}

Of particular interest in many studies of UDGs thus far has been their positioning within the stellar mass -- halo mass relationship of galaxies. In Figure \ref{fig:smhm}, we show the observed UDGs in stellar mass -- halo mass space in comparison to the two simulations. HESTIA simulated galaxies tend to lie above the observationally established stellar mass -- halo mass relationships of both \citet{Brook2014} and \citet{Danieli2023}, suggesting that they over-produce stars in their dark matter halos (e.g., via gas over-cooling). NIHAO simulated galaxies largely follow both relationships. This will include any UDGs that have formed in NIHAO (see also \citealp{Gannon2023}). To aid discussion, we colour two regions that are $>0.5$dex away from the stellar mass -- halo mass relationship of \citet{Brook2014} as overbright galaxies (cyan; i.e., where galaxies have more stellar mass than expected given their total halo mass) and as failed galaxies (olive; i.e., where galaxies have less stellar mass than expected given their total halo mass). UDGs are joined to the halo that best reproduces their dynamical mass (see Section \ref{sec:methods}) via a dotted grey line. The only difference for the plotted UDGs is their observed stellar mass (red) vs the stellar mass of their simulated best-fitting dark matter halo (blue point to which they are joined). The collection of UDGs on the left-hand side of the NIHAO plot have all been assigned to the lowest mass halo from the simulation, which did not produce any stars (i.e., a dark halo). As such, they have infinite $F_{\star, {\rm NIHAO}}$~values. 16/22 UDGs matched to NIHAO and 10/22 matched to HESTIA have halo masses $\ge 0.5$dex above their simulated counterparts and reside in the ``overbright galaxies'' region. This can also be seen by the large number of UDGs with high $F_\star$~values in Table \ref{tab:summary}. In contrast, only 3/22 galaxies matched to NIHAO and 5/22 matched to HESTIA have low $F_\star$~values and reside in the region of ``failed galaxies''. 

Unlike the evidence of many observational studies, our matching exercise would suggest that a large fraction of UDGs cannot be ``failed galaxies'' but are instead the opposite - dark matter halos with an abundance of stars, many more than are expected for their dark matter halo mass. We stress that current observations clearly demonstrate that such a high fraction of UDGs residing in low-mass dark matter halos is not the case. We refer the reader to Forbes \& Gannon (submitted) for a full discussion as to why UDG halo mass estimates (and in particular those coming from GC counts) are to be believed observationally. Further, we refer the reader to \citet[][fig. 11 and 12 and related discussion]{Zaritsky2023} for an argument based on UDGs' structural properties that they likely reside in dark matter halos that are more massive than dwarf galaxies of similar stellar masses. 

Based on the arguments presented in Forbes \& Gannon (submitted), we use the halo mass estimates from the established GC number -- halo mass relationship of \citet{Burkert2020} as the most robust measure of their halo mass. We compare these observed halo masses to their matched halo masses from the simulations in the next section.

\subsection{Observed Halo Masses vs. Simulated Halo Masses} \label{sec:hm_comparison}
\begin{figure*}
    \centering
    \includegraphics[width=0.95\textwidth]{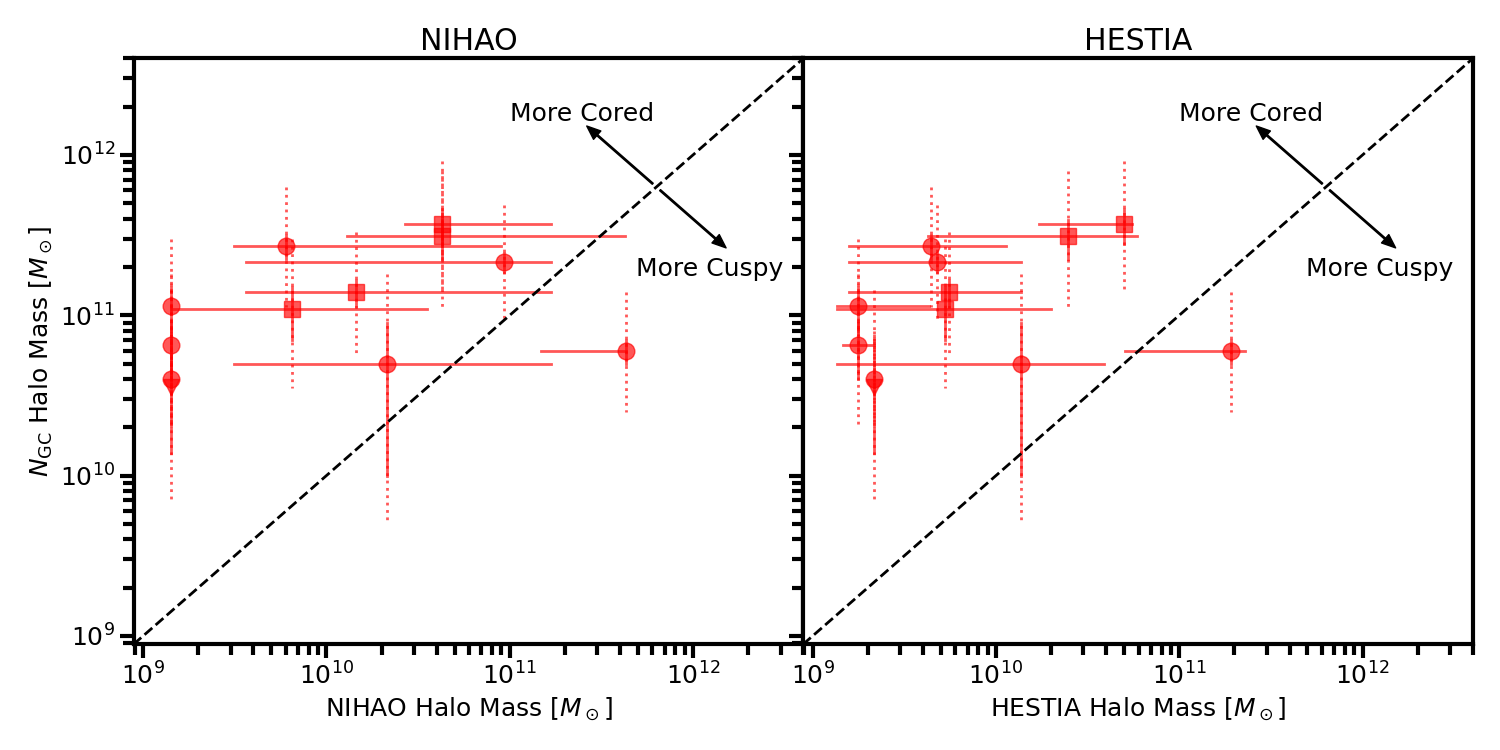}
    \caption{Observed halo mass based on UDG GC number \textit{vs.} matched halo mass in the simulation from the UDG's dynamical mass. Markers follow Figure \ref{fig:mass_radius}.} \textit{Left:} We show the NIHAO simulated halo masses that UDGs have been matched to. \textit{Right:} We show the HESTIA simulated halo masses that UDGs have been matched to. Observations are restricted to those with more than 5 GCs to ensure they have a robust halo mass estimate based on their GC number. In the y direction, uncertainties based on their GC count are plotted in solid red, with dotted extensions representing the addition of an assumed 0.3 dex scatter in the \citet{Burkert2020} relationship. A 1:1 line is shown as the black dashed line diagonally crossing the figure. Arrows are shown in the top right, indicating how the simulated halos would need to change to reproduce the observations. Observed UDG halo masses are frequently much greater than the haloes they are matched to in the simulations. Observed UDGs reside in fundamentally different dark matter halos to those that are being simulated. Frequently, these UDGs would require a dark matter halo with a larger core than is being simulated (even for NIHAO, which does produce dark matter cores).
    \label{fig:hmcompare}
\end{figure*}

Figure \ref{fig:hmcompare} compares the total halo mass inferred from observed UDG GC systems with the halo mass assigned to the same UDGs based on simulations. We exclude observed UDGs with less than 5 GCs to ensure we have an accurate estimate of their total halo mass. Uncertainties are included based on both the scatter in the relationship and the uncertainty in their GC counts and represent an upper limit for their halo mass uncertainty. Based on the GC number -- halo mass relationship, we take the halo masses from these GC counts (i.e., the y-axis) to be that which is observed for these UDGs. In general, a UDG's `true' halo mass from observations is much more massive than those from the NIHAO and HESTIA simulations (at least for those with $\ge5$GCs). \textit{Observed UDGs reside in fundamentally different dark matter halos to those that are being simulated.}

We take this statement to be particularly interesting since, by definition based on our methods, the observed dynamical masses for these UDGs agree with the mass enclosed within the same radius of the halo they have been matched to in the simulations. That is, many currently observed UDG dynamical masses are able to be reproduced by simulations, however, their total halo masses are not. We stress that this statement is true in Figure \ref{fig:hmcompare} even after considering both the uncertainty on these UDGs' GC numbers as listed in \citet{Gannon2024}, combined with an assumed 0.3 dex uncertainty in the GC number -- halo mass relationship as derived by \citet{Burkert2020} and the uncertainty of our matching methods. We discuss the possible solutions to this interesting puzzle below. 

\subsection{Solving the Tension} \label{sec:solutions}
Below we consider a few reasons, from both the observational and simulated perspectives, as to how observed UDGs and simulated UDGs can have similar enclosed dynamical masses while exhibiting $\sim1$dex different total halo masses.\\

\subsubsection{Observational Solutions} \label{sec:stripping}
To solve this dilemma observationally, measured dynamical masses need to underestimate the total dynamical mass within their half-light radii. The implication would be that the `true' mass within their half-light radius is larger than what is being inferred and, once this effect is accounted for, they will be matched to higher mass halos in the simulations, which will better correspond to their observed total halo masses. There are a number of ways this could be possible:

\begin{enumerate}
    \item \textit{Rotation}: Recent works of \citet{Chilingarian2019}, \citet{Buttitta2025}, \citet{Khim2025} and Levitsky et al (submitted) have found that some UDGs rotate, an effect not previously measurable in integrated measurements such as those presented in \citet{Gannon2022}. While integrated measurements naturally account for rotation along the line of sight \citep{Courteau2014}, they will not account for ``true rotation'' as an inclination correction would be needed. This correction would increase their measured dynamical masses, helping to match observed UDGs to higher mass halos from the simulations. The UDG definition is biased to face on objects (see e.g., \citealp{Li2023} or \citealp{Pfeffer2024}), and so this may represent a significant increase to their measured stellar velocity dispersions. It is worth noting however, that UDG stellar velocity dispersions largely follow the established stellar mass -- stellar velocity dispersion relationship \citep{Gannon2021, Toloba2023}, so any significant correction to their velocity dispersions (e.g., an increase of a factor $>2$) would cause inconsistencies elsewhere in our understanding of these UDGs (e.g., massively increase their already dark matter dominated nature within 1$R_{\rm e}$). 

    \item \textit{Environmental Processes:} NIHAO/HESTIA simulate UDGs in low-density environments while our comparison sample of UDGs is biased to higher-density cluster environments. Simulations such as Romulus \citep{Tremmel2020, Wright2021} suggest that field UDGs and cluster UDGs may form via separate formation pathways. The implication of the disparate dominant formation pathways for UDGs in low and high density environments in large volume simulations may be that simulated cluster UDGs have a halo profile of largely different characteristics to those in the field. As such, environmental processes may alter the central dynamical masses of observed UDGs, causing high mass halos to have lower dynamical masses in clusters than in the field. Likely, this would be due to tidally stripped dark matter. While other tidal processes, such as tidal heating, have also been proposed to form UDGs in clusters (e.g., \citealp{Carleton2019}), to reconcile the simulations we have examined with observations, an alteration of the halo profile is required. 
    
    In the case of tidally stripping dark matter, we note that for many UDGs, this tidal stripping would need to be strong to explain the difference between the dynamical mass of the massive simulated halos and the observed dynamical mass. Frequently, it would require a $>90\%$ decrease in dynamical mass. UDGs in clusters do not necessarily show the tidal features suggestive of the strong stripping that would be required to largely change their dynamical masses \citep{Mowla2017}. Further, any tidal stripping would have to occur without removing the UDGs' GC systems, from which we infer their large halo masses. Finally, we note that the cluster UDGs in Figure \ref{fig:mass_radius} exhibit, on average, higher dynamical masses than their group counterparts, which would not be expected if they were strongly tidally stripped. Resolving our tension by invoking tidal processes and the bias of our study to simulations of low-density environments would thus require simulations of cluster UDGs to exhibit strong tidal stripping, despite the observed UDGs in clusters they are being compared to exhibiting very little evidence for even mild tidal stripping. On current evidence, as a solution for the entire population, we suggest that environmental processes may be a little contrived.

    \item \textit{Mass Estimation Formula:} An assumption of our work is that the formula of \citet{Wolf2010} accurately reproduces the dynamical mass within the half-light radius for our UDGs. Given UDGs are amongst the most extreme galaxies at their stellar mass, it is possible this assumption is poor (see e.g., \citealp{Sarrato2025} for an example where the formula may underestimate masses in simulations). Obviously, if our dynamical masses are not accurately estimated, any inference drawn from them will be flawed.. In the point above, we have covered the assumption that these are dispersion-supported systems. \citet{Wolf2010} also makes the assumption of a relatively flat velocity dispersion profile near the half-light radius. Currently, there is no observational evidence that this is not the case, and the resolved velocity profile that was measured for DF44 is relatively flat \citep{vanDokkum2019b}. \citet{Wolf2010} also assumes that the galaxies are in dynamical equilibrium. While many of the observed UDGs are in dense clusters, making it possible they are currently being disrupted and are not in equilibrium, they do not tend to show signs of tidal disruption. There is also some assumption of spherical symmetry in the formula of \citet{Wolf2010}. Currently, there is a bias to UDGs with spectroscopy having higher axis ratios (lower ellipticities) than samples from imaging surveys \citep{Gannon2024}, making it likely that this assumption is more valid for our observed sample than that of the general UDG population. Finally, while the \citet{Wolf2010} formula was originally derived to maximise accuracy for halos with varying anisotropy, \citet{Errani2018} posit that, once variations in halo profile shape are included in these calculations, a slightly different mass estimator at a larger radius more accurately reproduces true masses. Here, we have used \citet{Wolf2010} due to its widespread use in the literature.  
    
    \item \textit{Velocity Anisotropy:} It is also possible that, despite the \citet{Wolf2010} formula being optimized to still produce accurate dynamical masses in the case of velocity anisotropy, there exists a significant enough velocity anisotropy within our UDGs such that their line of sight velocity dispersions poorly represent their total dynamical support. Similar to the addition of rotation, this would cause inconsistencies elsewhere in our understanding of UDGs. It would also require an explanation as to why the majority of currently observed UDGs have this anisotropy. Finally, the one UDG for which an isotropy can be inferred, DF44, has a slight preference to an isotropic (i.e., not anisotropic) orbital distribution \citep{vanDokkum2019b}. 
\end{enumerate}

One of the largest issues with observational solutions to the issue is the requirement to increase UDG dynamical masses to cause a better consistency between their simulated and observed halo masses. GC-rich UDGs are already amongst the most dark matter-dominated galaxies at their stellar mass \citep{Toloba2018, vanDokkum2019b, Gannon2020, Gannon2021}. Any increase in their observed dynamical masses via one of the above biases would result in them becoming \textit{even more} extreme in their central dark matter content. \\

\subsubsection{Simulated Solutions} 
\begin{figure}
    \centering
    \includegraphics[width=0.5\textwidth]{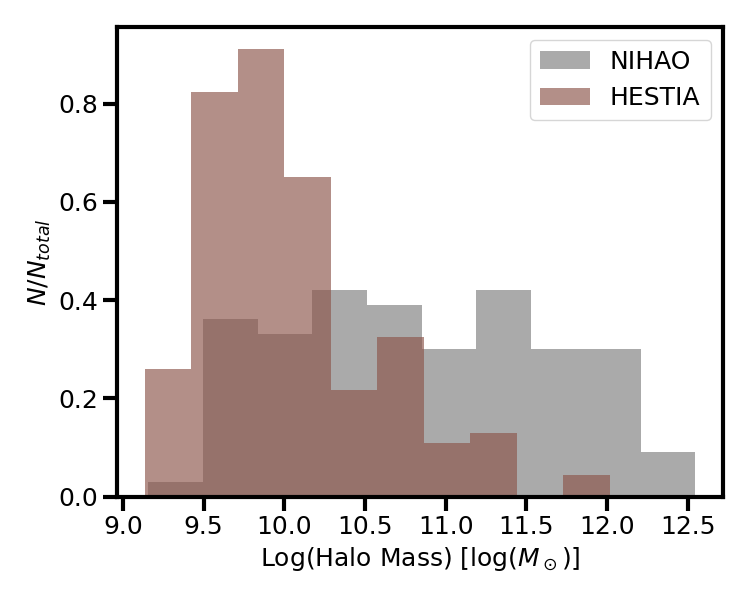}
    \caption{Normalised histograms of the total halo mass of the comparison halos from the simulations. NIHAO is plotted in grey and is distributed in a flat manner with halo mass. HESTIA is plotted in brown and has comparatively more low mass halos than high mass halos. In any given cosmological volume, there are more low-mass halos than high-mass halos (e.g., the HESTIA simulations) however, the NIHAO simulations are not reflective of cosmological halo abundances and display a flat distribution.  }
    \label{fig:halo_histograms}
\end{figure}

To solve this dilemma by adjusting the simulations, the dark matter halo profiles of the simulated UDGs must be incorrect. The implication being that a halo profile is required that has the same dynamical mass that the halos currently have in these two simulations, but significantly more dark matter at large radii, resulting in a more massive dark matter halo. We discuss some possible causes for this below: 

\begin{enumerate}
    \item \textit{Excluding Gas Content:} Our choice to exclude gas mass when making the comparison inevitably biases our results. However, including gas content will cause lower mass halos to create larger dynamical masses, resulting in even lower mass halos being matched to our UDGs. As such, our choice to exclude gas mass when making the matching causes us to present the tension in its most charitable form. Any inclusion of gas mass would create a larger difference between observations and simulations.   

    \item \textit{Greater Numbers of Low Mass Halos:} There is clearly a region of Figure \ref{fig:mass_radius} where NIHAO halos of mass $\sim10^{10} M_\odot$ and $\sim10^{11} M_\odot$ produce similar dynamical masses at a given radius. A greater discussion of halo profiles producing similar dynamical masses can be found in \citet{Gannon2021} or \citet[][see their fig. 6]{McQuinn2022}. Given a cosmological volume, there will exist more low-mass halos than high-mass halos, a result of $\Lambda$CDM that is true for both the Universe and cosmological simulations of it. The net result of these two effects, high mass halos producing similar dynamical masses to low mass halos and there being generally more low mass halos, would lead our matching scheme to be more likely to assign a low-mass halo to our UDGs than a high-mass one. However, we can rule this out as a cause for the discrepancy. In the case of the NIHAO simulations, they are by construction, not reflective of cosmological halo abundances and have a uniform selection with halo mass (as seen in Figure \ref{fig:halo_histograms}). For HESTIA, while it does exhibit more low mass halos than high mass halos, these halos do not vary in shape with mass and exhibit a cusp at all halo masses. As such, it is not possible for there to be overlap in dynamical mass for halos of largely different total mass. Therefore, we can rule out the greater numbers of low mass halos as a cause for the discrepancy in our study. Furthermore, our conclusions remain after the inclusion of uncertainties based on the minimum and maximum halo masses that pass through our dynamical mass measurements, which further helps mitigate this issue. 
    
    \item \textit{Poor Stellar Mass Reproduction:} Recent findings, such as \citet{Watkins2025}, have found that the light distributions of galaxies from the NEWHORIZON simulations are systematically different from observations. This is certainly the case for both NIHAO and HESTIA. In NIHAO, the vast majority of galaxies in the stellar mass range of $10^7 - 10^9 M_{\odot}$ are UDGs, with very few `normal' dwarf galaxies \citep{Jiang2019}. In HESTIA, dark matter halos overproduce stars, as is clearly evident in Figure \ref{fig:smhm}. As such, the distribution of stellar masses within the centre of their dark matter halos is likely very different to that of the observed UDGs, which will affect the dynamical masses. It is unlikely that this will have a sufficient effect to resolve our problem. Most UDGs are dark matter dominated, and the stellar mass distribution of a given galaxy is a relatively low fraction of the dynamical mass within its half-light radius (of order 10\%). 
    
    \item \textit{Larger Dark Matter Cores:} While NIHAO does produce dark matter cores (usually of size $\sim1 R_{\rm e}$), what is less clear is that it produces dark matter cores of sufficient size. Having a larger dark matter core would lower the central dynamical masses of massive halos, resulting in UDG dynamical masses being matched to higher mass halos. Given that UDGs are amongst the most diffuse galaxies known in a mass-follows-light understanding of dark matter halos, it seems possible that they inhabit the most diffuse dark matter halos. More specifically, \citet{Forbes2024} demonstrated that cores of the maximum extent expected for a \citet{Read2016} halo profile ($\sim 2.75 \times R_{\rm e}$) can reproduce both the dynamical masses and their massive dark matter halos. As UDGs have larger half-light radii than normal dwarf galaxies, defining core size by galaxy half-light radius will result in much larger dark matter cores ($>5$kpc) than are found in normal dwarf galaxies. It would be left as an outstanding challenge for simulations to produce these extended cores while simultaneously being able to produce the far less diffuse dark matter halos of normal dwarf galaxies. To be specific, it would be a requirement of simulations to self-consistently produce very large dark matter cores and normal cuspy halos (and everything in between) for dwarf galaxies of the same stellar mass. The lack of this diversity of dark matter halos in simulations is similar to the established `Diversity of Rotation Curves' problem for dwarfs \citep{Oman2015, Sales2022} and to the diversity of UDG concentrations proposed by \citet{Kravtsov2024}. 

    % \item \textit{Halo Profile Variance:} While NIHAO is able to produce both cored and cuspy dark matter halos, it does not tend to both at the same halo mass \citep{diCintio2014}. In HESTIA, dark matter halos tend to be relatively cuspy at all halo masses (see Figure \ref{fig:mass_radius}). To reproduce the dynamical masses currently being observed in many UDGs these simulations will need to create both cuspy (to produce normal galaxies) and cored (to produce UDGs) dark matter halos at fixed halo mass.

\end{enumerate}

Ultimately, we believe the most viable explanation is that the shape of the dark matter profile for UDGs in simulations is fundamentally different to the shape of the dark matter halos of observed UDGs. That is, to place observed UDGs in more massive simulated haloes, while retaining the same dynamical mass, one has to create cores of much larger size than those that are obtained by either simulation. As an example, the NIHAO simulations form cores of typical size $\sim1 R_{\rm e}$. \citet{diCintio2014} finds that core creation is most efficient in halos with a logarithmic stellar to halo mass ratio $M_\star / M_{\rm Halo} \approx -2~ \mathrm{to}~ -3$ which is exactly the range expected for currently observed UDGs, suggesting that formation should be extremely efficient. A core of larger physical extent is then required, with \citet{Forbes2024} finding cores of size $\sim2.75\times R_{\rm e}$ best reproduce current UDG observations. Our findings are echoed by the recent work of \citet{ManceraPina2025}, which found systematically lower dark matter halo concentrations for dwarf galaxies than are being simulated. \textit{We suggest there is a need for a greater diversity in simulated dwarf galaxy dark matter halo profiles as our solution to the problem presented herein.}

% we believe that an observationally motivated solution to the discrepancy between observed UDG masses and those simulated, as presented in our paper, will pose more issues to the field. That is not the case for proposed solutions that suggest there should be a larger diversity of dwarf galaxy profile shapes in simulations than is currently simulated. 

% Due to this, we prefer the need for a greater diversity in simulated dwarf galaxy dark matter halo profiles as our solution to the problem presented herein. 

\section{Conclusions} \label{sec:conclusions}
In this work, we have compared observed UDGs to those from the NIHAO and HESTIA simulations. We started by matching observed dynamical masses to the best-fitting dark matter halo from each simulation and comparing the total stellar mass observed to that of the galaxy formed in the simulation. Our approach differs from the one presented in \citet{Forbes2024} whereby here we use simulated halos rather than the idealised analytical comparison made there. We then compared our matched halo masses in simulations to the observed halo masses for those UDGs from GC counts. Our main conclusions are as follows:

\begin{itemize}
    \item Current observed UDG dynamical masses are matched to simulated haloes of less mass than a simple comparison of their observed stellar masses to the stellar mass -- halo mass relationship would suggest. Further, the simulated galaxies that reside in the halos to which the observations are matched tend to have less stellar mass than is observed for the UDGs. This implies a disconnect between the dark matter halos in simulations that are creating UDG-like dynamical masses and the dark matter halos expected from the observational properties of UDGs. 

    % Current observed UDG dynamical masses, once matched to simulated haloes, imply halos that are less massive than a simple comparison to the stellar mass -- halo mass relationship would suggest. Further, the simulated galaxies that reside in the halos that best match UDG dynamical masses tend to have much less stellar mass than the observed UDGs.

    % \item We place 4 galaxies (three of them UDGs) with independent halo mass measurements on the established GC number -- halo mass relationship. We find an excellent agreement, even down to relatively low number GC systems. We conclude that GC-rich UDGs can have accurate halo masses estimated using this relationship.

    \item We compare the halo masses observed in UDGs using the GC number -- halo mass relationship to those inferred from our NIHAO/HESTIA matching, finding a large offset. In general, UDGs are observed to have much higher halo masses than their dynamical masses would suggest if compared to simulations. This presents a puzzle as it is not clear why simulations would be able to produce halos of similar dynamical (central) mass without simultaneously producing the right halo (total) mass. 

    \item We discuss some possible solutions to the puzzle from both an observational and simulated perspective. We find the most plausible solution is that there is a need for a greater diversity in halo profile shapes for dwarf galaxies than is currently being simulated. Namely, as dwarf galaxies are observed to be everything from small and compact to large and diffuse at a similar stellar mass, there is growing observational evidence that their dark matter halos may be similarly diverse (i.e., from cuspy and concentrated to cored and diffuse at fixed total halo mass). Here we provide evidence for the need to place UDGs in dark matter halos with cores of much larger size than are usually produced, one extreme of the above halo profile diversity. Reproducing this diversity is a key requirement of future dwarf galaxy simulations. We stress that this reproduction must be done self-consistently within a cosmological simulation. i.e., it is insufficient to demonstrate that a simulation can produce a dark matter halo with a large radius for a dwarf galaxy without also producing a cuspy dark matter halo at the same stellar mass. 
\end{itemize}

\section*{Acknowledgements}
We thank the anonymous referee for their detailed and constructive reading of our work. JSG completed a significant portion of this paper during a FINCA visiting programme in Feb, 2025. He is grateful for their support. DAF, JPB, WJC thank the ARC for financial support via DP220101863 and DP250101673. ADC kindly thanks the Centre for Astrophysics and Supercomputing (CAS) for the financial support during her visit to Swinburne University, through their Women Visiting Fellowship program, and the Spanish Ministerio de Ciencia, Innovacion y Univerasidades through grant CNS2023-144669, programa Consolidacion Investigadora.

%%%%%%%%%%%%%%%%%%%%%%%%%%%%%%%%%%%%%%%%%%%%%%%%%%
\section*{Data Availability}
This work makes use of a publicly available catalogue of UDG spectroscopic properties available \href{https://github.com/gannonjs/Published_Data/tree/main/UDG_Spectroscopic_Data}{here}. The full references for the catalogue are \citet{mcconnachie2012, vanDokkum2015, Beasley2016, Martin2016, Yagi2016, MartinezDelgado2016, vanDokkum2016, vanDokkum2017, Karachentsev2017, vanDokkum2018, Toloba2018, Gu2018, Lim2018, RuizLara2018, Alabi2018, FerreMateu2018, Forbes2018, MartinNavarro2019, Chilingarian2019, Fensch2019, Danieli2019, vanDokkum2019b, torrealba2019, Iodice2020, Collins2020, Muller2020, Gannon2020, Lim2020, Muller2021, Forbes2021, Shen2021, Ji2021, Huang2021, Gannon2021, Gannon2022, Mihos2022, Danieli2022, Villaume2022, Webb2022, Saifollahi2022, Janssens2022, Gannon2023, FerreMateu2023, Toloba2023, Iodice2023, Shen2023, Janssens2024, Gannon2024, Buttitta2025}. Simulated data will be made available upon reasonable request with the corresponding author.



\bibliographystyle{mnras}
\bibliography{bibliography} % if your bibtex file is called example.bib

%\begin{thebibliography}{99}
%\end{thebibliography}

%%%%%%%%%%%%%%%%%%%%%%%%%%%%%%%%%%%%%%%%%%%%%%%%%%

%%%%%%%%%%%%%%%%% APPENDICES %%%%%%%%%%%%%%%%%%%%%

% \appendix

% \section{Some extra material}
% Possibly move the HESTIA comparison here if it interrupts the flow of the paper. 

%%%%%%%%%%%%%%%%%%%%%%%%%%%%%%%%%%%%%%%%%%%%%%%%%%

% Don't change these lines
\bsp	% typesetting comment
\label{lastpage}
\end{document}